\documentclass[
reprint,
superscriptaddress,
 amsmath,amssymb,
 aps,
]{revtex4-2}

\usepackage{graphicx}
\usepackage{dcolumn}
\usepackage{bm}

\begin{document}

\preprint{APS/123-QED}

\title{Lessons from Transforming Second-Year Honors Physics Lab}

\author{Danny Doucette}
\email{danny.doucette@pitt.edu}
\affiliation{Department of Physics and Astronomy, University of Pittsburgh, Pittsburgh, Pennsylvania 15260, USA}

\author{Brian D'Urso}
\affiliation{Department of Physics, Montana State University, Bozeman, Montana 59717, USA}

\author{Chandralekha Singh}
\affiliation{Department of Physics and Astronomy, University of Pittsburgh, Pittsburgh, Pennsylvania 15260, USA}

\date{\today}

\begin{abstract}
New technology like the Arduino microcontroller platform presents an opportunity to transform Beyond the First Year (BFY) physics labs to better prepare physics students for work in research labs and beyond. The flexibility, low cost, and power of these devices provides an attractive way for students to learn to use and master research-grade instrumentation. Therefore, we introduced new technology, including Arduino Due microcontroller boards, to a second-year honors physics lab in order to provide improved learning experiences for students. This transformation was implemented in three lab modules and focused on diminishing the black box nature of the traditional labs while encouraging students to engage in troubleshooting. The importance of troubleshooting was made evident to students by the instructor emphasizing it as an inevitable and central part of experimentation. This lab transformation also required that students perform work that was `above and beyond' the scope of the assigned experimental work for part of the course credit. While the technological aspects of the transformation were received well by a majority of students, our observations during the initial implementation suggested a need for some modifications to instructional practices in order to improve the learning and experiences for all students. In particular, we find that many students can benefit from additional scaffolding in order to complete `above and beyond' work. Similarly, we find that students in general, and underrepresented students such as women in particular, may need thoughtful intervention from the instructor, e.g., in order to avoid becoming isolated when the lab work is designed for pair work. Otherwise, some students may be left to work alone with a disproportionate work-load if students choose their own partners. With these lessons taken into account, recent student experiences in the transformed lab were notably improved.
\end{abstract}

\maketitle

\section{Introduction}

Labs designed for physics majors are essential for preparing the next generation of physicists. These labs should provide physics students with opportunities to learn to think like a physicist in a lab context and teach them essential skills that will be useful for both academic and non-academic careers. However, in typical labs encountered by physics majors, including labs beyond the first year, one typically finds complex, expensive equipment that bear little resemblance to what would be used in an academic research lab or on a job site. While undergraduate research experiences may provide a bridge between undergraduate classes and research labs~\cite{HunterBecomingAScientist}, they often pose the same difficulty of either requiring or at least preferring skills and familiarity with equipment that isn't typically found in lab courses~\cite{LinnUREs}.

Thus, there is a need to transform instructional labs in order to better equip physics majors with skills they can use in undergraduate research experiences, graduate school research, or non-academic jobs. The creative and purposeful use of computers and/or low-cost electronics is a popular and successful approach for retooling physics labs~\cite{BeichnerActiveLearningSpaces,DoriBelcherTechActiveLearning,MastersOscilloscope, Galvez5QMExps, BeckLowCostElectronics, GeorgeComputerSupport,  SpaldingCoValuing, SpaldingSinglePhoton,SobhanzadehLaboratorials}, and is part of the strategy we adopted to transform honors physics lab with three new lab modules at our institution.

At the University of Pittsburgh, first-year physical science students (including physics majors) take a separate 2-credit lab course after their first introductory college physics course rather than taking introductory labs as part of, or in parallel with, their first introductory physics classes. Thus, the first physics lab that physics majors take is either the regular introductory lab in the spring semester of their first year of studies or the honors lab, discussed here, that is offered in the fall semester of their second year of studies. This honors lab requires a certain minimum grade in the introductory physics lecture course and has a reputation of requiring intense work, with two three-hour lab sessions weekly. Enrollment is typically between 10 and 16 students. Given the advanced preparation of many of its students, the topics of investigation, and the scheduling during the second year of studies, the honors lab may best be compared with Beyond the First Year (BFY) labs at other institutions even though it is the first college physics lab taken by the students enrolled in it. Approximately 30\% of our physics majors take this honors lab, while the remaining 70\% take a regular, non-honors, lab which also enrolls chemistry, engineering, and other majors. 87\% of the students in the honors lab are physics majors, while only 20\% of students in the regular, non-honors, lab choose to major in physics. 

The process of transforming a physics lab has been well-explicated in the physics education literature~\cite{ZwicklAdvLab, HolmesOperationalizingAAPT}. The first step was to identify goals by consulting with stakeholders such as faculty members and referring to documents such as the AAPT lab guidelines~\cite{AAPTlab}. Second, we focused on designing new lab work, procedures, and apparatus consistent with the feedback obtained. Finally, we evaluated the transformation, which involved three new Arduino-based lab modules as well as additional features such as an `above and beyond' task associated with each lab.

Two overarching goals led our efforts: helping students learn to think like physicists, and teaching essential undergraduate-level research techniques to students. These goals led to two guiding principles in the design of the three new modules in our transformation. First, that equipment be research-grade, used across multiple experiments, and not merely a `black box'. Second, that students be able to troubleshoot equipment, explore variables and parameters in the experiment, and work collaboratively. To address the first set of principles, an Arduino-based digital test instrument was developed and deployed. All software developed for this transformation is open source and available online~\cite{Pythics}, and additional specifications for how the system is set up are available on request. To address the second set of principles, changes were made to the way the course was run, including the adoption of an `above and beyond' requirement for lab work.

The perspective of the cognitive apprenticeship model~\cite{CognitiveApprenticeship}, which was useful in this context of transforming the lab, proposes that learning is effective when the criteria of good performance are modeled explicitly and then students are provided coaching and scaffolding support to learn important skills before they can successfully practice those skills independently. We believe that scaffolding new skills is essential for students in lab courses. Both troubleshooting and the `above and beyond' work will be productive for students only if they are provided with the guidance, scaffolding and support they need in order to be successful in the new, transformed lab. To evaluate the impact of the three new Arduino-based modules and `above and beyond' task associated with the transformed lab, we analyzed student work, evaluated student attitudes toward experimental science using the E-CLASS survey, and conducted observations and interviews of students enrolled in the lab.

\section{New Lab Modules}

\begin{figure}
\includegraphics[width=\columnwidth]{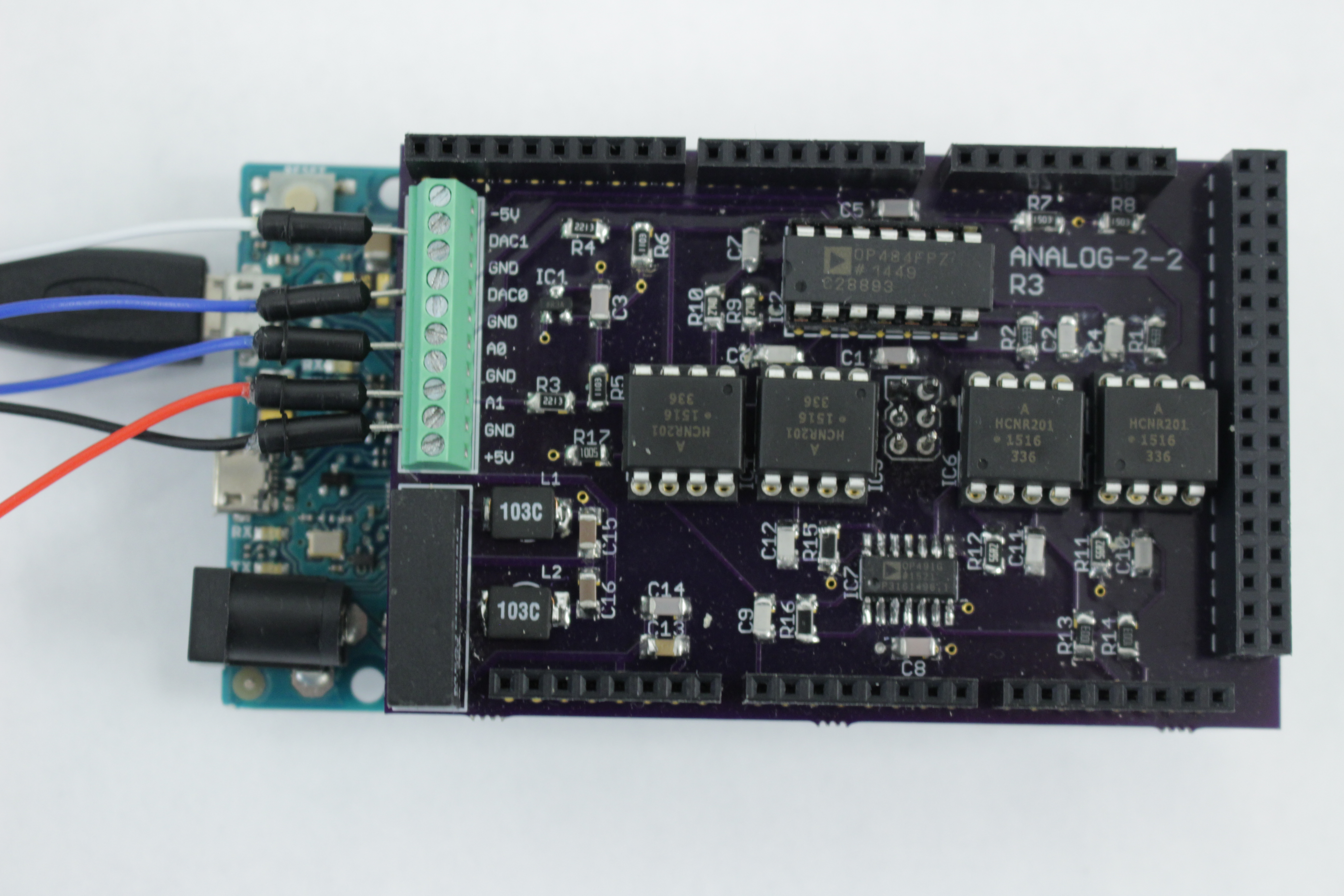}
\caption{The Arduino Due and custom-built shield used for input and output and analog electrical signals. The Arduino (green, underneath) connects to the computer by USB cable. The shield (black) sits on top of the Arduino and connects with the experiment via jumper cables.\label{ArduinoShield}}
\end{figure}

The simplicity, low-cost, ubiquity, and capabilities of the Arduino Due make it an effective instrument for the physics lab~\cite{BouquetArduino}. While previous work~\cite{HsuArduino,GaleriuArduino,LavelleArduino,GopalakrishnanArduino} has primarily focused on using the Arduino microcontroller boards for exciting one-off investigations, we report here on a flexible and powerful system that brings research-grade electronics to some of experiments students perform in the second-year honors teaching lab. In combination with a simple shield (a custom-built board which interfaces with the Arduino Due, see Fig.~\ref{ArduinoShield}), a breadboard, and a computer, the Arduino-based system is able to replicate the capabilities of a variety of traditional lab equipment. Data are transferred from the Arduino to a computer, where open-source software~\cite{Pythics} performs analysis to replicate the functions of an oscilloscope, a synthesizer, a lock-in amplifier, a spectrum analyzer, and a network analyzer. The focus is on helping students learn to deeply understand and debug issues related to the software of one piece of equipment, rather than poring over the details of many complicated devices~\cite{DeVoreLockinAJP, DeVoreLockinPRPER}. Encouraging and teaching troubleshooting skills was a major goal of the lab transformation.

While the lab seeks to help students discern value in both computer-based data acquisition systems and electronic test equipment such as multimeters and oscilloscopes (both of which are used frequently in the second half of the course), there are several advantages to helping students learn to use the Arduino-based instrument first, rather than using a collection of separate lock-in amplifiers, spectrum analyzers, network analyzers, and other expensive test equipment for lab work. The breadboard ecosystem allows students to use inexpensive, easily-replaced components, and the Arduino-based system itself is inexpensive in comparison with lock-in amplifiers and network analyzers that are available on the market today.

Students are guided and encouraged to view the Arduino-based instrumentation as a tool that they can dig into and understand: they are taught some basic Python in the first weeks of the course and encouraged to explore and edit the data analysis code that runs on the computer, and to do the same with the data acquisition code that runs on the Arduino. Furthermore, students typically co-enroll, or take in a subsequent semester, physics courses in scientific programming and digital electronics which use Python and the Arduino, respectively, so that -- at least in principle -- students should be able to understand and work with the Arduino-based instrumentation at a deeper level by the time they are ready to begin undergraduate work in a faculty research lab. A major goal of introducing these new lab modules that use the Arduino-based instrumentation was to encourage students to move beyond thinking of their data acquisition system as a black box.

The Arduino Due itself is built around a powerful microcontroller of a type that is commonly used in academic and industrial research labs (an ARM Cortex-M3, with a 280 kHz sample rate and 12-bit analog input/output bit depth), so that expertise with Arduino systems can be ported directly into research-grade lab work.

Students, in general, responded positively to the introduction of the Arduino in the three new modules of this honors lab course, expressing enthusiasm about their lab work and indicating excitement about coming to the lab course. Some students reported that they were able to take their lab skills to research labs and immediately start working on research. More than 50\% of students, and 100\% of students in the most recent offering of the course, elected to perform one or more Arduino-based experiments when given the choice (see Table~\ref{Experiments}.)

Below, we describe the three new experiments that were designed to use the Arduino-based instrument. They supplant experiments that relied on equipment for which parts are no longer available, or for which traditional equipment was not up to the task of adequately performing standard analyses. The experiments are also designed so that students are slowly introduced to new capabilities of their instrumentation and the features of the experimental apparatus, which allows them to learn to think systematically about how their instruments may be misbehaving and troubleshoot more effectively. All three new experiments were carefully written to meet our overarching goals of helping students learn to think like physicists and teaching students techniques useful for experimental research.

\begin{table*}
\caption{Experiments available to students in this course. All students complete Test Measurements, and they must complete at least one other from the leftmost column. Star indicates a new experiment that uses the Arduino-based instrumentation.\label{Experiments}}
\begin{ruledtabular}
\begin{tabular*}{\textwidth}{lll}
Test Measurements* & Single Photon Interference & Noise Fundamentals\\
RLC Circuits* & Electron Diffraction & Nuclear Magnetic Resonance\\
Acoustical Resonance* & Photoelectric Effect & Chaotic Circuit\\
Acoustical Gas Thermometer & Black-Body Radiation & Muon Lifetime\\
Electron Spin Resonance & Microwave Optics & Radiation Detection\\
\end{tabular*}
\end{ruledtabular}
\end{table*}

\textbf{Test Measurements:} After a four-week introduction to lab procedures and software, including an introduction to Python programming and data analysis techniques for physics, students work through this first lab to learn how to use a variety of test equipment. This equipment includes an oscilloscope, function generator, multimeter, spectrum analyzer, lock-in amplifier, and network analyzer, all of which run on the Arduino-based instrument. After some introductory measurements, students pass signals from a conventional or Arduino-based function generator through an RC circuit and qualitatively measure the amplitude and phase response at a given input frequency using an oscilloscope. Then they use a spectrum analyzer and lock-in amplifier to examine the response more quantitatively. Next, the network analyzer function is used to measure the amplitude and phase response over a wide range of frequencies and to show how RC circuits can be used as low/high-pass filters. Finally, students are given freedom to explore the Fourier decomposition of a square wave or repeat the analysis using an RL circuit.

This lab is used to introduce the Arduino-based instruments to students. The use of a conceptually-simple circuit makes it easier to scaffold student understanding of both the functioning of the new instruments as well as the signal processing that is being performed. The Arduino-based lock-in amplifier, spectrum analyzer, and network analyzer make it possible to have all students complete this lab at the same time, without needing to share costly equipment. This lab introduces students to a number of important fundamental principles about electronics that they will revisit in their advanced-level electronics lab course, including equipment such as breadboards, instrumentation, and methods for troubleshooting. After this experiment, students rotate among a selection of two-week experiments, completing four, for which there may be only one set of apparatus available. These experiments are listed in Table~\ref{Experiments} (but may not all be operational at any given time). The following two labs are included in the rotation.

\textbf{RLC Circuits:} In this two-week lab, students first examine the transient response in an RLC circuit using the synthesizer and oscilloscope. After that, they drive the circuit with a sine wave and investigate the steady-state response using the oscilloscope, spectrum analyzer, and lock-in amplifier. They also use the network analyzer to see how this circuit behaves around resonance. Some emphasis is placed on measuring and understanding the phase response of the circuit and comparing the response to that of a harmonic oscillator. The goal is to combine elements of electronics (which will, as with the Test Measurements lab, be built upon in the advanced electronics lab course) with the interesting physics of resonance.

While the Test Measurements lab takes students through a scaffolded, step-by-step procedure, this experiment is more open-ended. After introducing the theory and illustrating the circuit to build, the instructions give students a few specific issues to investigate, while giving students plenty of freedom about which tools to use, the specific parameters to use, and so forth.

\textbf{Acoustic Resonance:} In this two-week lab, students examine acoustic resonant modes in a wooden box using a speaker and microphone. Students use the oscilloscope, and eventually the lock-in amplifier and network analyzer, to measure the frequencies of modes, as well as the phase response. Finally, the students move the microphone around the box in order to map out a resonant mode in physical space. This serves as a useful contrast to the RLC circuits lab, as it brings the concept of resonance into physical space for students to explore. As with the RLC Circuits experiment, this lab is also quite open-ended, with students given plenty of scope to explore and address questions they find interesting on the way through the procedure. While the RLC circuit has a simple, single resonance, the acoustical cavity has much increased complexity due to the presence of many resonances with varying spatial distributions.

\section{Student Learning}

Alongside the new instrumentation, several elements of the course were changed with the goals of the transformation in mind. Lab handouts were rewritten from a highly-structure format to a more informal discussion of topics to investigate. The grading scheme was changed to emphasize the importance of maintaining a useful lab notebook~\cite{StanleyLabNotebooks}. The change to Arduino-based instruments softened the black box nature of instrumentation that is typical in physics labs, which necessitated more time spent upfront helping students learn how the Arduino-based instrument works. 

A benefit from this approach to understanding the instrumentation is that it makes troubleshooting a central, and repeatedly-emphasized, element of the lab course. The instructor of the course motivated the transformation by focusing on the importance of troubleshooting and how it can help students learn to think like physicists. Students need guidance in order to learn explicitly how to troubleshoot their apparatus, and scaffolding needs to be provided for learning troubleshooting techniques~\cite{DeVoreLockinAJP,DounasFrazerInstructorTroubleshooting}. Troubleshooting was explicitly taught in two ways. First, in the Test Measurements module, students were stepped through the experiment in such a way that they could learn how to test the ways in which individual electronic components affected an electrical signal. They used probes to make measurements at several points as a signal passed a circuit, and were asked to explain how and why the signal changed at each point. A common and important form of troubleshooting in this lab was tracing signals through electric circuits on the breadboard in order to diagnose circuit wiring difficulties. Second, lab instructors made a point of helping students learn to diagnose and address issues with their experiments in a supportive way throughout the course. One way that instructors provided this assistance was by suggesting specific measurements that students could make in order to produce results that would be helpful in diagnosing the issue they were encountering.

Another change was the requirement that 20\% of students' lab reports discuss explorations `above and beyond' the scope of the work assigned in the lab handout. Students each wrote a complete traditional lab report for their final experiment, which was graded using a rubric that focused on the clarity of the description of the lab work and the correctness of the analysis. The `above and beyond' requirement aligns well with the goal of getting students to think like physicists, as this independent exploration is an exemplary opportunity for them to develop their curiosity, their skills in designing and conducting experiments, and their physics identities~\cite{CarloneJohnsonIdentity,HazariIdentity,KalenderIdentity1,KalenderIdentity2}. For example, a student working on an interference experiment did a calculation to show that there was (usually) only one photon present in the device at a time. Another example was when a student noticed an unexpected behavior in one of the graphs, and followed up with some insightful analysis of how the Arduino might have a non-negligible internal resistance or inductance.

\begin{figure}
\includegraphics[width=\columnwidth]{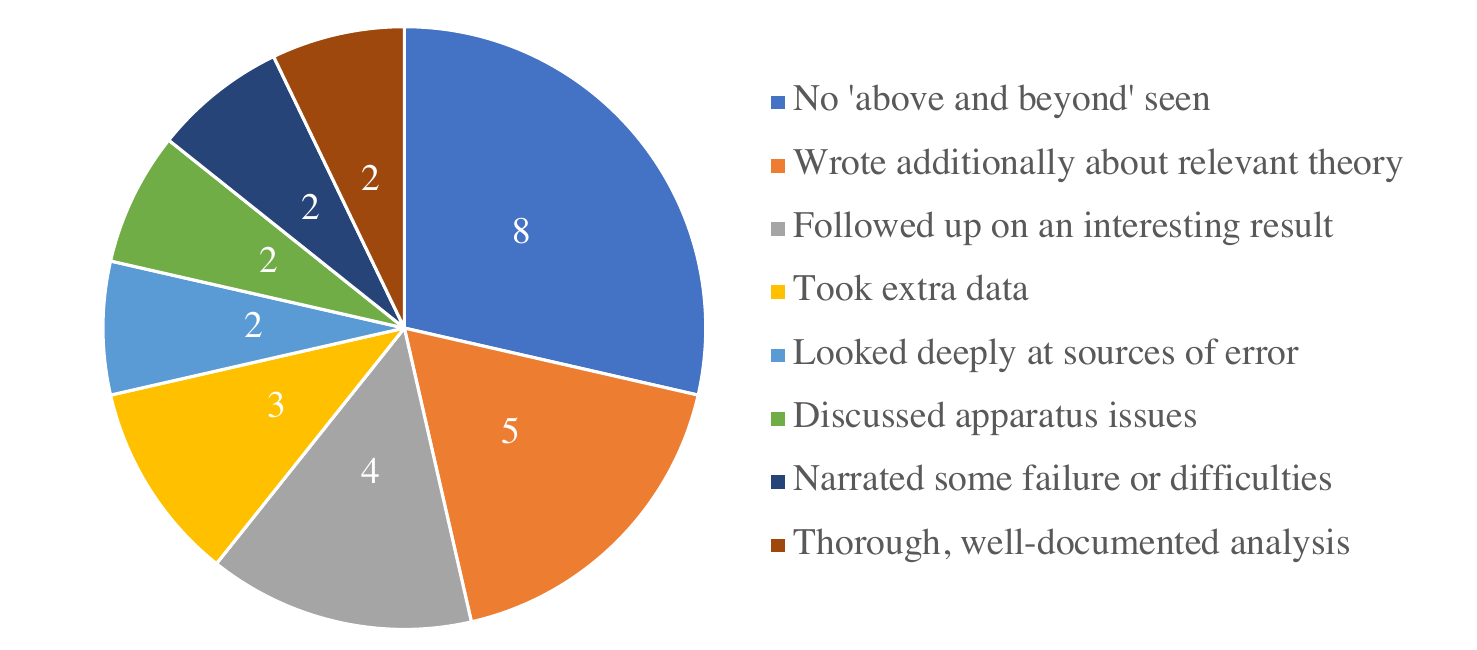}
\caption{Types of `above and beyond' work done by students in the transformed lab. Numbers in parentheses indicate the number of student reports in each category. All of the work except for ``No `above and beyond' seen'' were awarded full points for the 20\% of the lab report grade allocated to the `above and beyond' requirement.\label{AboveAndBeyond}}
\end{figure}

Using a generative coding scheme~\cite{OteroGettingStarted} developed and validated by the authors, an analysis of 28 lab reports from one semester of the transformed course in which students completed more than one lab report found a large variety of `above and beyond' work being done by students, as shown in Fig.~\ref{AboveAndBeyond}. Some students chose to investigate and write about the theoretical side of their results, relying on their textbook and internet research to demonstrate a deeper understanding of the underlying theory. Some reports showed students following up on a result that caught their eye by attempting to understand what might have caused that result. Other approaches included taking additional data to extend the range of their investigation, performing a deeper dive into the error analysis, writing about potential improvements that could be made to the apparatus, and providing a narrative-style description of one part of the experiment that did not work out as expected. Most impressive were the lab reports in which students made a substantial, meaningful, and well-explained extension to their lab work. In these cases, the students truly went `above and beyond', demonstrating independent investigation skills and aptitudes that show they are well-prepared for a research setting. 

However, the limited amount of scaffolding provided for the `above and beyond' work made it difficult for a sizable number of students to excel in this aspect of the lab course. These students, whose work shows up in the approximately 30\% of Fig. \ref{AboveAndBeyond} categorized as ``no `Above and Beyond' seen'', may have needed more guidance and feedback, and may have benefited from seeing examples of what this type of work is, or could have benefited from some additional support regarding the type of work that would be reasonable or acceptable. In the most recent offering of the lab course, examples of past `above and beyond' work were provided, and this was viewed as quite useful by the students.

\begin{figure*}
\begin{tabular}{ll}
\includegraphics[width=0.5555\textwidth]{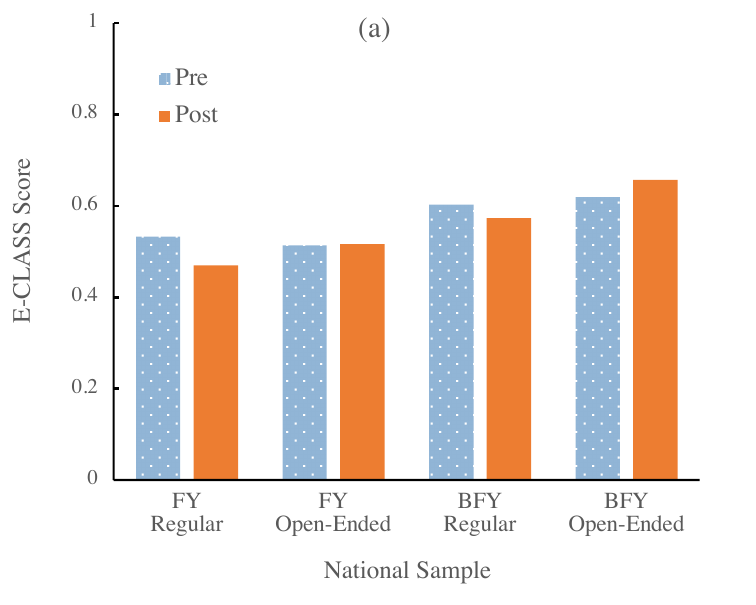}
&
\includegraphics[width=0.4444\textwidth]{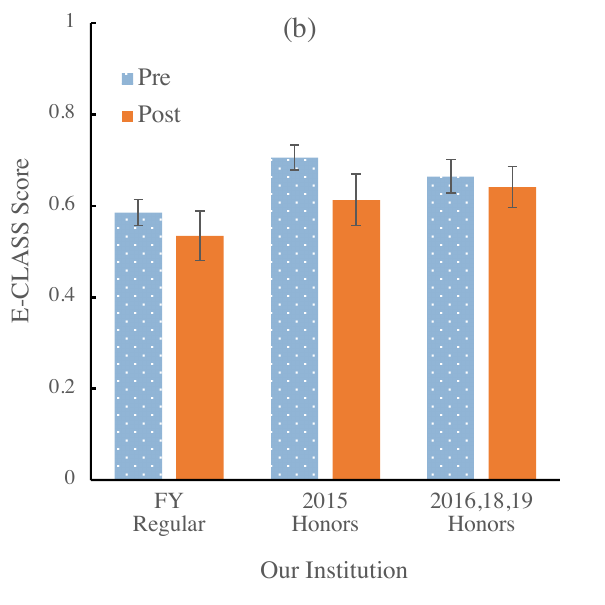}
\end{tabular}
\caption{(a) Pre (blue) and post (orange) E-CLASS scores for a national sample of students separated into First-Year (FY) and Beyond First Year (BFY) labs, and whether students primarily do regular, traditional lab work ($n = 3487$ for FY and $279$ for BFY) or open-ended experiments ($n = 592$ for FY and $553$ for BFY), from Wilcox and Lewandowski~\cite{WilcoxOpenEnded}. (b) Scores from our institution include a regular, non-honors, introductory lab that is taken by physical science majors, including physics students who do not take the honors lab ($n=49$); the honors lab before transformation (only data from 2015 is available, $n=12$); and the honors lab for three offerings after the transformation (2016, 2018, and 2019, $n=33$). Error bars indicate standard error in mean E-CLASS scores.\label{ECLASS}}
\end{figure*}

In order to evaluate how student attitudes toward experimental physics have changed as a result of the transformation, the E-CLASS survey was given to students at the beginning (pre) and end (post) of the semester. E-CLASS scores indicate the extent to which participants hold expert-like views about experimental physics, including strategies, habits of mind, and attitudes as measured through 30 items with a 5-point Likert scale~\cite{ECLASSValidation}. In keeping with precedent, we express scores from $-1$ (novice-like) to $+1$ (expert-like) and focus on students' own responses, and not how they believe an expert would respond, which the E-CLASS survey also captures. Fig.~\ref{ECLASS}(a) indicates scores from a nationally-representative sample~\cite{WilcoxOpenEnded} and Fig.~\ref{ECLASS}(b) shows scores from our university. For the honors physics lab, our results suggest a larger decrease in E-CLASS scores the year prior to the transformation, 2015, (effect-size given by Cohen's $d=0.67$, t-test $p=0.08$~\cite{CohenD}) and a smaller decrease during post-transformation years, ($d=0.10$, $p=0.31$). Although the number of students in these two samples is small, and so the results are not statistically significant, the fact that the effect size for the decrease given by Cohen's $d$ goes from $0.67$ to $0.10$ is encouraging. The E-CLASS scores for the honors lab are similar to those for BFY courses in the national sample ($d=0.02, p=0.10$; compare BFY in Fig.~\ref{ECLASS}(a) with Honors in Fig.\ref{ECLASS}(b)). The difference as measured by Cohen's $d$ in this case is ``small''~\cite{CohenD}. Analysis of the national sample in Fig.~\ref{ECLASS}(a) conducted by Wilcox and Lewandowski indicated that students in regular, traditional labs typically see a decrease in E-CLASS scores, while labs that include open-ended work (like our transformed labs) produce scores that are unchanged or increase slightly between pre and post~\cite{WilcoxOpenEnded}. 


Moreover, the selective nature of the honors lab is apparent in Fig. \ref{ECLASS}, as E-CLASS pre scores for our honors lab are consistently higher than pre scores in both first-year (FY) labs from the national sample~\cite{WilcoxOpenEnded} (t-test $p<0.001$) as well as our regular, non-honors, lab ($p<0.001$). In addition to being selective and potentially attracting students with high levels of prior preparation, the majority of the students who enroll in the honors lab are male physics majors. Most female physics majors choose to enroll in the regular, non-honors, introductory lab, in which physics students account for 20\% of the enrollment. However, even though students in the honors lab may be considered a privileged group overall compared with the regular introductory lab, that doesn't mean they all have sufficient prior experience, e.g., with experimental techniques, and don't need to be supported in their learning in the lab context.

\section{Student Experiences}

The physics lab can be a culturally-rich, low-stakes environment for students to develop useful research skills and stimulate their interest in physics. Formative lab experiences, like learning to use the Arduino, can help some students to develop an interest in physics and come to see themselves as physicists or scientists. From the beginning of the transformation, students generally responded positively. Most students who were introduced to the Arduino boards quickly became familiar with the platform, and were soon able to put the device to use in a variety of creative ways. Some students explicitly expressed enthusiasm for the transformation and indicated that they looked forward to coming to the lab. Other students subsequently sought out research opportunities and reported that they were able to immediately begin working productively in a research lab using skills and understandings they had developed in class. However, for some students who may not have received appropriate support, lab experiences can negatively impact their physics identity trajectory~\cite{DoucetteHermione}. In order to understand the impact of this lab transformation on student experiences, we conducted 12 hours of observations in the lab and interviewed 7 students who took the transformed lab.

The lab environment generally seemed positive both before and after the transformation, with some male students cracking jokes (often involving experiments and troubleshooting) and good relationships between instructors and students. Students were free to work alone or in pairs, and nearly all chose to work with a partner. During years with an odd number of students, this typically meant that one student would work alone. After the transformation, the instructor repeatedly emphasized the importance of learning to troubleshoot apparatus: being patient, learning from mistakes, and being deliberate and methodical in learning how to find and diagnose potential issues from unexpected results. While troubleshooting is indeed a skill that we need to help students learn, it can be frustrating and overwhelming if a student does not have a partner or they don't feel comfortable asking for help, since students may not have the self-confidence and prior skills to effectively carry out the troubleshooting task without appropriate guidance and scaffolding support. In particular, one difficulty with our initial approach was that it put the onus on students to ask the instructor for assistance if they were struggling. 

For students who end up working alone, and for those with lower levels of self-efficacy, it sometimes meant that students didn't feel comfortable asking for help. Based on discussions with students, we found that this affected two students in the first years of the lab transformation. A male student who did not have a partner because he was the `odd one out' in his year described the lab as very difficult and felt very negatively about his experiences. He struggled to complete the experimental work and write up a lab report by himself. This is not surprising since the amount of lab work was designed to be shared by two students. A female student from a different year of the lab transformation, who described this class as ``totally awesome'' in principle, explained how being the only woman in the lab course meant that her male classmates ``kind of paired off'', leaving her to work alone. As an unsupported solo student, she reported that the course ``was just really hard... there isn't that much guidance, and that can make it really difficult, especially because I was doing it without a partner.'' She described struggling with the equipment and having difficulty managing the troubleshooting while working alone, saying, ``If I don't know what I'm doing, it's hard for me to say what help I need because I don't know where I'm going wrong in this." She added, ``You start off and you have the oscilloscopes and you have to hook up a bunch of stuff with cables, which I had never done before, and I didn't know what all these cables were and where you're supposed to get them, and it was a ton of just learning how to do measurements.'' Frustrated by the struggle of working alone on a lab designed for two students, her experiences led her to question whether she belonged in physics labs, saying ``I just didn't feel like I was supposed to be there at a certain point.''

While both the male and the female student found working alone to be difficult, being left alone to do all the work had a more detrimental effect on the female student. Perhaps because of a lower sense of belonging and a lower level of self-efficacy~\cite{CauseForAlarm,KalenderSelfEfficacy}, the female student dropped out of the course, enrolling in the regular introductory lab instead, while the male student in the same situation was frustrated by being left alone but managed to persist and complete the class. 

We note that, since it may be more difficult for underrepresented students such as women to find a partner to work with in labs such as these, it is important that instructors be deliberate and thoughtful in ensuring that students are not left to work alone if they are not adequately supported to succeed while doing so. After identifying this challenge, we took steps to ensure that subsequent offerings of the honors lab would allow students the opportunity to work in groups of 3 if necessary, that the instructor would form groups if anyone was being left out, and that the instructor would take initiative to visit students working on labs in order to provide more support to students who need it, whether they ask for it explicitly or not.

In the most recent offering of the course, the female students who enrolled in the course found partners and no students dropped out because of needing to work alone. Moreover, two female students who worked together in the most recent offering of the transformed lab described their experience in the class as a very positive one. They appreciated that the lab instructor, aware of the challenges faced by students in this lab, frequently checked in with them and was willing to answer their questions quietly in a corner of the room where they wouldn't be overheard by other students. For students with a lower sense of belonging, it can be particularly valuable to be able to ask questions without fear of being judged by their peers. Similarly, a female student who worked with a male partner described the lab as being a good course, challenging but fair.

\section{Discussion and Summary}

Based on our experiences, we offer the following advice for instructors of similar labs who wish to ensure their students are being given equal learning opportunities and not being overly burdened when they have students work in groups:

\begin{enumerate}
    \item Ensure all students have an opportunity to benefit from working with a partner~\cite{SinghPeersTest}, even if they might initially prefer to work alone.
    \item Assign students to groups if needed, respecting that students may benefit from choosing their own partners when they work in pairs~\cite{AzmitiaFriends}. For groups of more than two, be careful to avoid marooning underrepresented students (e.g., a woman working with two men) and attend to the roles students assume in their group work~\cite{HellerBook,DoucetteHermione}. As the female student who dropped the transformed lab after being left to work alone in the earlier implementation of the lab explained, ``Making the teachers assign partners, as uncool as that is, would definitely help''. 
    \item Rotating groups occasionally (e.g., every few weeks) can help students break out of bad work habits and learn to work with different partners~\cite{HellerBook}.
    \item Check in with students frequently. Seek to make it normal to ask questions and seek guidance without fear of being judged, and take advantage of the opportunity to provide guidance to students out of earshot of their peers where they can feel more comfortable (this may be particularly beneficial for underrepresented students).
\end{enumerate}

Labs that target physics majors have the potential to provide students the opportunity to develop experimental skills and expertise with research-grade equipment and help them learn to think like a physicist. The Arduino-based instrument described here is a flexible and powerful device that helps to meet this opportunity. Meanwhile, changes in the structure of the course such as supporting the development of troubleshooting skills, scaffolding `above and beyond' work, and adopting open-ended skills-based work could help students to develop essential scientific skills for future research in academic and non-academic settings~\cite{KirwanTaskAnalysis, ClarkTaskAnalysis}.

While the spectrum of types of `above and beyond' work shown in Fig.~\ref{AboveAndBeyond} is impressive, we are in the process of both improving the overall quality of this type of work and helping students who do not undertake `above and beyond' work for all their lab reports. We may think of this challenge as seeking to balance the innovative aspect of student work with the need for students to practice skills to improve their efficiency in the work they do in the lab~\cite{BransfordSchwartz}. In other words, we need to ensure that the struggle students experience in the lab is productive~\cite{KapurProductiveStruggle} and that students are not frustrated with the open-ended nature of the troubleshooting and `above and beyond' tasks. Moving forward, we plan to provide improved support and guidance for students as they learn to troubleshoot their apparatus, such as by playing out simulated problems in a coordinated way.

Past research suggests that active learning (of which lab work is an example) can contribute to decreasing the performance `gap' between overrepresented and underrepresented groups of students~\cite{LorenzoCrouchMazur}. However, further investigations suggest that it is the implementation of active learning that is the critical factor in determining whether active learning is able to close these performance gaps. In particular, one study~\cite{KarimEBAEgender} showed that while all students learned more in evidence-based active learning classes, the performance gap between men and women increased from pre to post test. This is the main reason we focused explicitly on the experiences of individual students in our evaluation of this transformed honors lab via individual interviews and lab observations. More generally, our findings support the claim that failing to attend to the needs of traditionally-underrepresented groups of students in the lab risks perpetuating inequities in physics~\cite{DoucetteHermione}. For the honors lab, which is taken by many of our physics majors, this first college experimental physics experience is critical for students who may already struggle to see themselves as physicists~\cite{CarloneJohnsonIdentity}. Therefore, we conducted individual interviews with a subset of students in order to highlight the importance of attending to the needs of underrepresented students even if there are very few such students. The struggles of these types of students will not be captured by aggregate data such as the E-CLASS scores in Fig.~\ref{ECLASS}.

The transformation of a second-year honors lab curriculum offers an exciting opportunity to increase the relevance, accessibility, and quality of an essential physics learning experience. It is important for such transformations to be done in a way that ensures that students are adequately supported and carefully accounts for the complete processes of anticipated skill development. By doing this, new lab courses such as the one introduced here can provide a venue for all students to develop both positive physics identity and valuable research-ready lab skills.

\begin{acknowledgments}
We thank Russell Clark, Istvan Danko, Bob Devaty, Gurudev Dutt and the faculty at the University of Pittsburgh for support and informative conversations and we acknowledge the invaluable insight provided by our undergraduate interview participants in this investigation. This material is based upon work supported by the National Science Foundation under Grant No. 1504874 and 1834463.
\end{acknowledgments}

\bibliography{bibfile}

\begin{thebibliography}{45}%
\makeatletter
\providecommand \@ifxundefined [1]{%
 \@ifx{#1\undefined}
}%
\providecommand \@ifnum [1]{%
 \ifnum #1\expandafter \@firstoftwo
 \else \expandafter \@secondoftwo
 \fi
}%
\providecommand \@ifx [1]{%
 \ifx #1\expandafter \@firstoftwo
 \else \expandafter \@secondoftwo
 \fi
}%
\providecommand \natexlab [1]{#1}%
\providecommand \enquote  [1]{``#1''}%
\providecommand \bibnamefont  [1]{#1}%
\providecommand \bibfnamefont [1]{#1}%
\providecommand \citenamefont [1]{#1}%
\providecommand \href@noop [0]{\@secondoftwo}%
\providecommand \href [0]{\begingroup \@sanitize@url \@href}%
\providecommand \@href[1]{\@@startlink{#1}\@@href}%
\providecommand \@@href[1]{\endgroup#1\@@endlink}%
\providecommand \@sanitize@url [0]{\catcode `\\12\catcode `\$12\catcode
  `\&12\catcode `\#12\catcode `\^12\catcode `\_12\catcode `\%12\relax}%
\providecommand \@@startlink[1]{}%
\providecommand \@@endlink[0]{}%
\providecommand \url  [0]{\begingroup\@sanitize@url \@url }%
\providecommand \@url [1]{\endgroup\@href {#1}{\urlprefix }}%
\providecommand \urlprefix  [0]{URL }%
\providecommand \Eprint [0]{\href }%
\providecommand \doibase [0]{https://doi.org/}%
\providecommand \selectlanguage [0]{\@gobble}%
\providecommand \bibinfo  [0]{\@secondoftwo}%
\providecommand \bibfield  [0]{\@secondoftwo}%
\providecommand \translation [1]{[#1]}%
\providecommand \BibitemOpen [0]{}%
\providecommand \bibitemStop [0]{}%
\providecommand \bibitemNoStop [0]{.\EOS\space}%
\providecommand \EOS [0]{\spacefactor3000\relax}%
\providecommand \BibitemShut  [1]{\csname bibitem#1\endcsname}%
\let\auto@bib@innerbib\@empty
\bibitem [{\citenamefont {Hunter}\ \emph {et~al.}(2007)\citenamefont {Hunter},
  \citenamefont {Laursen},\ and\ \citenamefont
  {Seymour}}]{HunterBecomingAScientist}%
  \BibitemOpen
  \bibfield  {author} {\bibinfo {author} {\bibfnamefont {A.~B.}\ \bibnamefont
  {Hunter}}, \bibinfo {author} {\bibfnamefont {S.~L.}\ \bibnamefont
  {Laursen}},\ and\ \bibinfo {author} {\bibfnamefont {E.}~\bibnamefont
  {Seymour}},\ }\bibfield  {title} {\bibinfo {title} {Becoming a scientist: The
  role of undergraduate research in students' cognitive, personal, and
  professional development},\ }\href {https://doi.org/10.1002/sce.20173}
  {\bibfield  {journal} {\bibinfo  {journal} {Science Education}\ }\textbf
  {\bibinfo {volume} {91}},\ \bibinfo {pages} {36} (\bibinfo {year}
  {2007})}\BibitemShut {NoStop}%
\bibitem [{\citenamefont {Linn}\ \emph {et~al.}(2015)\citenamefont {Linn},
  \citenamefont {Palmer}, \citenamefont {Baranger}, \citenamefont {Gerard},\
  and\ \citenamefont {Stone}}]{LinnUREs}%
  \BibitemOpen
  \bibfield  {author} {\bibinfo {author} {\bibfnamefont {M.~C.}\ \bibnamefont
  {Linn}}, \bibinfo {author} {\bibfnamefont {E.}~\bibnamefont {Palmer}},
  \bibinfo {author} {\bibfnamefont {A.}~\bibnamefont {Baranger}}, \bibinfo
  {author} {\bibfnamefont {E.}~\bibnamefont {Gerard}},\ and\ \bibinfo {author}
  {\bibfnamefont {E.}~\bibnamefont {Stone}},\ }\bibfield  {title} {\bibinfo
  {title} {Undergraduate research experiences: Impacts and opportunities},\
  }\href {https://doi.org/10.1126/science.1261757} {\bibfield  {journal}
  {\bibinfo  {journal} {Science}\ }\textbf {\bibinfo {volume} {347}},\ \bibinfo
  {pages} {627} (\bibinfo {year} {2015})}\BibitemShut {NoStop}%
\bibitem [{\citenamefont {Beichner}(2014)}]{BeichnerActiveLearningSpaces}%
  \BibitemOpen
  \bibfield  {author} {\bibinfo {author} {\bibfnamefont {R.}~\bibnamefont
  {Beichner}},\ }\bibfield  {title} {\bibinfo {title} {History and evolution of
  active learning spaces},\ }in\ \href@noop {} {\emph {\bibinfo {booktitle}
  {Active Learning Spaces}}}\ (\bibinfo  {publisher} {Jossey-Bass},\ \bibinfo
  {year} {2014})\ pp.\ \bibinfo {pages} {9--16}\BibitemShut {NoStop}%
\bibitem [{\citenamefont {Dori}\ and\ \citenamefont
  {Belcher}(2005)}]{DoriBelcherTechActiveLearning}%
  \BibitemOpen
  \bibfield  {author} {\bibinfo {author} {\bibfnamefont {Y.~J.}\ \bibnamefont
  {Dori}}\ and\ \bibinfo {author} {\bibfnamefont {J.}~\bibnamefont {Belcher}},\
  }\bibfield  {title} {\bibinfo {title} {How does technology-enabled active
  learning affect undergraduate students' understanding of electromagnetism
  concepts?},\ }\href@noop {} {\bibfield  {journal} {\bibinfo  {journal} {The
  Journal of the Learning Sciences}\ }\textbf {\bibinfo {volume} {14}},\
  \bibinfo {pages} {243} (\bibinfo {year} {2005})}\BibitemShut {NoStop}%
\bibitem [{\citenamefont {Masters}\ and\ \citenamefont
  {Miers}(1997)}]{MastersOscilloscope}%
  \BibitemOpen
  \bibfield  {author} {\bibinfo {author} {\bibfnamefont {M.~F.}\ \bibnamefont
  {Masters}}\ and\ \bibinfo {author} {\bibfnamefont {R.~E.}\ \bibnamefont
  {Miers}},\ }\bibfield  {title} {\bibinfo {title} {Use of a digital
  oscilloscope as a spectrum analyzer in the undergraduate laboratory},\ }\href
  {https://doi.org/10.1119/1.18748} {\bibfield  {journal} {\bibinfo  {journal}
  {American Journal of Physics}\ }\textbf {\bibinfo {volume} {65}},\ \bibinfo
  {pages} {254} (\bibinfo {year} {1997})}\BibitemShut {NoStop}%
\bibitem [{\citenamefont {Galvez}\ \emph {et~al.}(2005)\citenamefont {Galvez},
  \citenamefont {Holbrow}, \citenamefont {Pysher}, \citenamefont {Martin},
  \citenamefont {Courtemanche}, \citenamefont {Heilig},\ and\ \citenamefont
  {Spencer}}]{Galvez5QMExps}%
  \BibitemOpen
  \bibfield  {author} {\bibinfo {author} {\bibfnamefont {E.~J.}\ \bibnamefont
  {Galvez}}, \bibinfo {author} {\bibfnamefont {C.~H.}\ \bibnamefont {Holbrow}},
  \bibinfo {author} {\bibfnamefont {M.~J.}\ \bibnamefont {Pysher}}, \bibinfo
  {author} {\bibfnamefont {J.~W.}\ \bibnamefont {Martin}}, \bibinfo {author}
  {\bibfnamefont {N.}~\bibnamefont {Courtemanche}}, \bibinfo {author}
  {\bibfnamefont {L.}~\bibnamefont {Heilig}},\ and\ \bibinfo {author}
  {\bibfnamefont {J.}~\bibnamefont {Spencer}},\ }\bibfield  {title} {\bibinfo
  {title} {Interference with correlated photons: Five quantum mechanics
  experiments for undergraduates},\ }\href {https://doi.org/10.1119/1.1796811}
  {\bibfield  {journal} {\bibinfo  {journal} {American Journal of Physics}\
  }\textbf {\bibinfo {volume} {73}},\ \bibinfo {pages} {127} (\bibinfo {year}
  {2005})}\BibitemShut {NoStop}%
\bibitem [{\citenamefont {Branning}\ \emph {et~al.}(2009)\citenamefont
  {Branning}, \citenamefont {Bhandari},\ and\ \citenamefont
  {Beck}}]{BeckLowCostElectronics}%
  \BibitemOpen
  \bibfield  {author} {\bibinfo {author} {\bibfnamefont {D.}~\bibnamefont
  {Branning}}, \bibinfo {author} {\bibfnamefont {S.}~\bibnamefont {Bhandari}},\
  and\ \bibinfo {author} {\bibfnamefont {M.}~\bibnamefont {Beck}},\ }\bibfield
  {title} {\bibinfo {title} {Low-cost coincidence-counting electronics for
  undergraduate quantum optics},\ }\href {https://doi.org/10.1119/1.3116803}
  {\bibfield  {journal} {\bibinfo  {journal} {American Journal of Physics}\
  }\textbf {\bibinfo {volume} {77}},\ \bibinfo {pages} {667} (\bibinfo {year}
  {2009})}\BibitemShut {NoStop}%
\bibitem [{\citenamefont {George}\ \emph {et~al.}(2013)\citenamefont {George},
  \citenamefont {Broadstock},\ and\ \citenamefont
  {Abad}}]{GeorgeComputerSupport}%
  \BibitemOpen
  \bibfield  {author} {\bibinfo {author} {\bibfnamefont {E.~A.}\ \bibnamefont
  {George}}, \bibinfo {author} {\bibfnamefont {M.~J.}\ \bibnamefont
  {Broadstock}},\ and\ \bibinfo {author} {\bibfnamefont {J.~V.}\ \bibnamefont
  {Abad}},\ }\bibfield  {title} {\bibinfo {title} {Learning energy, momentum,
  and conservation concepts with computer support in an undergraduate physics
  laboratory},\ }in\ \href@noop {} {\emph {\bibinfo {booktitle} {Fourth
  International Conference of the Learning Sciences}}}\ (\bibinfo {year}
  {2013})\ pp.\ \bibinfo {pages} {2--3}\BibitemShut {NoStop}%
\bibitem [{\citenamefont {Spalding}\ and\ \citenamefont
  {Mc{C}ann}(2016)}]{SpaldingCoValuing}%
  \BibitemOpen
  \bibfield  {author} {\bibinfo {author} {\bibfnamefont {G.~C.}\ \bibnamefont
  {Spalding}}\ and\ \bibinfo {author} {\bibfnamefont {L.~I.}\ \bibnamefont
  {Mc{C}ann}},\ }\bibfield  {title} {\bibinfo {title} {Co-valuing instructional
  laboratory course offerings},\ }in\ \href
  {http://www.enfusestem.org/wp-content/uploads/2016/03/2016-Endiuse-Paper-Spalding-McCann.pdf}
  {\emph {\bibinfo {booktitle} {Proceedings of the Symposium on Envisioning the
  Future of Undergraduate STEM Education}}}\ (\bibinfo {address} {Washington,
  DC},\ \bibinfo {year} {2016})\BibitemShut {NoStop}%
\bibitem [{\citenamefont {Aspden}\ \emph {et~al.}(2016)\citenamefont {Aspden},
  \citenamefont {Padgett},\ and\ \citenamefont
  {Spalding}}]{SpaldingSinglePhoton}%
  \BibitemOpen
  \bibfield  {author} {\bibinfo {author} {\bibfnamefont {R.~S.}\ \bibnamefont
  {Aspden}}, \bibinfo {author} {\bibfnamefont {M.~J.}\ \bibnamefont
  {Padgett}},\ and\ \bibinfo {author} {\bibfnamefont {G.~C.}\ \bibnamefont
  {Spalding}},\ }\bibfield  {title} {\bibinfo {title} {Video recording true
  single-photon double-slit interference},\ }\href
  {https://doi.org/10.1119/1.4955173} {\bibfield  {journal} {\bibinfo
  {journal} {American Journal of Physics}\ }\textbf {\bibinfo {volume} {84}},\
  \bibinfo {pages} {671} (\bibinfo {year} {2016})}\BibitemShut {NoStop}%
\bibitem [{\citenamefont {Sobhanzadeh}\ \emph {et~al.}(2017)\citenamefont
  {Sobhanzadeh}, \citenamefont {Kalman},\ and\ \citenamefont
  {Thompson}}]{SobhanzadehLaboratorials}%
  \BibitemOpen
  \bibfield  {author} {\bibinfo {author} {\bibfnamefont {M.}~\bibnamefont
  {Sobhanzadeh}}, \bibinfo {author} {\bibfnamefont {C.~S.}\ \bibnamefont
  {Kalman}},\ and\ \bibinfo {author} {\bibfnamefont {R.~I.}\ \bibnamefont
  {Thompson}},\ }\bibfield  {title} {\bibinfo {title} {Labatorials in
  introductory physics courses},\ }\href
  {https://doi.org/10.1088/1361-6404/aa8757} {\bibfield  {journal} {\bibinfo
  {journal} {European Journal of Physics}\ }\textbf {\bibinfo {volume} {38}},\
  \bibinfo {pages} {065702} (\bibinfo {year} {2017})}\BibitemShut {NoStop}%
\bibitem [{\citenamefont {Zwickl}\ \emph {et~al.}(2013)\citenamefont {Zwickl},
  \citenamefont {Finkelstein},\ and\ \citenamefont
  {Lewandowski}}]{ZwicklAdvLab}%
  \BibitemOpen
  \bibfield  {author} {\bibinfo {author} {\bibfnamefont {B.~M.}\ \bibnamefont
  {Zwickl}}, \bibinfo {author} {\bibfnamefont {N.}~\bibnamefont
  {Finkelstein}},\ and\ \bibinfo {author} {\bibfnamefont {H.~J.}\ \bibnamefont
  {Lewandowski}},\ }\bibfield  {title} {\bibinfo {title} {The process of
  transforming an advanced lab course: Goals, curriculum, and assessments},\
  }\href {https://doi.org/10.1119/1.4768890} {\bibfield  {journal} {\bibinfo
  {journal} {American Journal of Physics}\ }\textbf {\bibinfo {volume} {81}},\
  \bibinfo {pages} {63} (\bibinfo {year} {2013})}\BibitemShut {NoStop}%
\bibitem [{\citenamefont {Holmes}\ and\ \citenamefont
  {Smith}(2019)}]{HolmesOperationalizingAAPT}%
  \BibitemOpen
  \bibfield  {author} {\bibinfo {author} {\bibfnamefont {N.~G.}\ \bibnamefont
  {Holmes}}\ and\ \bibinfo {author} {\bibfnamefont {E.~M.}\ \bibnamefont
  {Smith}},\ }\bibfield  {title} {\bibinfo {title} {Operationalizing the {AAPT}
  learning goals for the lab},\ }\href {https://doi.org/10.1119/1.5098916}
  {\bibfield  {journal} {\bibinfo  {journal} {The Physics Teacher}\ }\textbf
  {\bibinfo {volume} {57}},\ \bibinfo {pages} {296} (\bibinfo {year}
  {2019})}\BibitemShut {NoStop}%
\bibitem [{\citenamefont {{AAPT Committee on Laboratories}}(2015)}]{AAPTlab}%
  \BibitemOpen
  \bibfield  {author} {\bibinfo {author} {\bibnamefont {{AAPT Committee on
  Laboratories}}},\ }\href@noop {} {\emph {\bibinfo {title} {AAPT
  Recommendations for the Undergraduate Physics Laboratory Curriculum}}}\
  (\bibinfo  {publisher} {AAPT},\ \bibinfo {year} {2015})\BibitemShut {NoStop}%
\bibitem [{\citenamefont {D'Urso}(2015)}]{Pythics}%
  \BibitemOpen
  \bibfield  {author} {\bibinfo {author} {\bibfnamefont {B.}~\bibnamefont
  {D'Urso}},\ }\href {https://github.com/dursobr/Pythics} {\bibinfo {title}
  {Pythics}} (\bibinfo {year} {2015}),\ \bibinfo {note}
  {https://github.com/dursobr/Pythics}\BibitemShut {NoStop}%
\bibitem [{\citenamefont {Collins}\ \emph {et~al.}(1991)\citenamefont
  {Collins}, \citenamefont {Brown},\ and\ \citenamefont
  {Holum}}]{CognitiveApprenticeship}%
  \BibitemOpen
  \bibfield  {author} {\bibinfo {author} {\bibfnamefont {A.}~\bibnamefont
  {Collins}}, \bibinfo {author} {\bibfnamefont {J.~S.}\ \bibnamefont {Brown}},\
  and\ \bibinfo {author} {\bibfnamefont {A.}~\bibnamefont {Holum}},\ }\bibfield
   {title} {\bibinfo {title} {Cognitive apprenticeship: Making thinking
  visible},\ }\href@noop {} {\bibfield  {journal} {\bibinfo  {journal}
  {American Educator}\ }\textbf {\bibinfo {volume} {15}},\ \bibinfo {pages} {6}
  (\bibinfo {year} {1991})}\BibitemShut {NoStop}%
\bibitem [{\citenamefont {Bouquet}\ \emph {et~al.}(2017)\citenamefont
  {Bouquet}, \citenamefont {Bobroff}, \citenamefont {Fuchs-Gallezot},\ and\
  \citenamefont {Maurines}}]{BouquetArduino}%
  \BibitemOpen
  \bibfield  {author} {\bibinfo {author} {\bibfnamefont {F.}~\bibnamefont
  {Bouquet}}, \bibinfo {author} {\bibfnamefont {J.}~\bibnamefont {Bobroff}},
  \bibinfo {author} {\bibfnamefont {M.}~\bibnamefont {Fuchs-Gallezot}},\ and\
  \bibinfo {author} {\bibfnamefont {L.}~\bibnamefont {Maurines}},\ }\bibfield
  {title} {\bibinfo {title} {Project-based physics labs using low-cost
  open-source hardware},\ }\href {https://doi.org/10.1119/1.4972043} {\bibfield
   {journal} {\bibinfo  {journal} {American Journal of Physics}\ }\textbf
  {\bibinfo {volume} {85}},\ \bibinfo {pages} {216} (\bibinfo {year}
  {2017})}\BibitemShut {NoStop}%
\bibitem [{\citenamefont {Hsu}\ \emph {et~al.}(2017)\citenamefont {Hsu},
  \citenamefont {Dhingra},\ and\ \citenamefont {D'Urso}}]{HsuArduino}%
  \BibitemOpen
  \bibfield  {author} {\bibinfo {author} {\bibfnamefont {J.-F.}\ \bibnamefont
  {Hsu}}, \bibinfo {author} {\bibfnamefont {S.}~\bibnamefont {Dhingra}},\ and\
  \bibinfo {author} {\bibfnamefont {B.}~\bibnamefont {D'Urso}},\ }\bibfield
  {title} {\bibinfo {title} {Design and construction of a cost-efficient
  {A}rduino-based mirror galvanometer system for scanning optical microscopy},\
  }\href {https://doi.org/10.1119/1.4972046} {\bibfield  {journal} {\bibinfo
  {journal} {American Journal of Physics}\ }\textbf {\bibinfo {volume} {85}},\
  \bibinfo {pages} {68} (\bibinfo {year} {2017})}\BibitemShut {NoStop}%
\bibitem [{\citenamefont {Galeriu}\ \emph {et~al.}(2015)\citenamefont
  {Galeriu}, \citenamefont {Letson},\ and\ \citenamefont
  {Esper}}]{GaleriuArduino}%
  \BibitemOpen
  \bibfield  {author} {\bibinfo {author} {\bibfnamefont {C.}~\bibnamefont
  {Galeriu}}, \bibinfo {author} {\bibfnamefont {C.}~\bibnamefont {Letson}},\
  and\ \bibinfo {author} {\bibfnamefont {G.}~\bibnamefont {Esper}},\ }\bibfield
   {title} {\bibinfo {title} {An {A}rduino investigation of the {RC} circuit},\
  }\href {https://doi.org/10.1119/1.4917435} {\bibfield  {journal} {\bibinfo
  {journal} {The Physics Teacher}\ }\textbf {\bibinfo {volume} {53}},\ \bibinfo
  {pages} {285} (\bibinfo {year} {2015})}\BibitemShut {NoStop}%
\bibitem [{\citenamefont {Lavelle}(2018)}]{LavelleArduino}%
  \BibitemOpen
  \bibfield  {author} {\bibinfo {author} {\bibfnamefont {C.~M.}\ \bibnamefont
  {Lavelle}},\ }\bibfield  {title} {\bibinfo {title} {Gamma ray spectroscopy
  with {Arduino UNO}},\ }\href {https://doi.org/10.1119/1.5026595} {\bibfield
  {journal} {\bibinfo  {journal} {American Journal of Physics}\ }\textbf
  {\bibinfo {volume} {86}},\ \bibinfo {pages} {384} (\bibinfo {year}
  {2018})}\BibitemShut {NoStop}%
\bibitem [{\citenamefont {Gopalakrishnan}\ and\ \citenamefont
  {G{\"u}hr}(2015)}]{GopalakrishnanArduino}%
  \BibitemOpen
  \bibfield  {author} {\bibinfo {author} {\bibfnamefont {M.}~\bibnamefont
  {Gopalakrishnan}}\ and\ \bibinfo {author} {\bibfnamefont {M.}~\bibnamefont
  {G{\"u}hr}},\ }\bibfield  {title} {\bibinfo {title} {A low-cost mirror mount
  control system for optics setups},\ }\href
  {https://doi.org/10.1119/1.4895343} {\bibfield  {journal} {\bibinfo
  {journal} {American Journal of Physics}\ }\textbf {\bibinfo {volume} {83}},\
  \bibinfo {pages} {186} (\bibinfo {year} {2015})}\BibitemShut {NoStop}%
\bibitem [{\citenamefont {DeVore}\ \emph
  {et~al.}(2016{\natexlab{a}})\citenamefont {DeVore}, \citenamefont {Gauthier},
  \citenamefont {Levy},\ and\ \citenamefont {Singh}}]{DeVoreLockinAJP}%
  \BibitemOpen
  \bibfield  {author} {\bibinfo {author} {\bibfnamefont {S.}~\bibnamefont
  {DeVore}}, \bibinfo {author} {\bibfnamefont {A.}~\bibnamefont {Gauthier}},
  \bibinfo {author} {\bibfnamefont {J.}~\bibnamefont {Levy}},\ and\ \bibinfo
  {author} {\bibfnamefont {C.}~\bibnamefont {Singh}},\ }\bibfield  {title}
  {\bibinfo {title} {Improving student understanding of lock-in amplifiers},\
  }\href {https://doi.org/10.1119/1.4934957} {\bibfield  {journal} {\bibinfo
  {journal} {American Journal of Physics}\ }\textbf {\bibinfo {volume} {84}},\
  \bibinfo {pages} {52} (\bibinfo {year} {2016}{\natexlab{a}})}\BibitemShut
  {NoStop}%
\bibitem [{\citenamefont {DeVore}\ \emph
  {et~al.}(2016{\natexlab{b}})\citenamefont {DeVore}, \citenamefont {Gauthier},
  \citenamefont {Levy},\ and\ \citenamefont {Singh}}]{DeVoreLockinPRPER}%
  \BibitemOpen
  \bibfield  {author} {\bibinfo {author} {\bibfnamefont {S.}~\bibnamefont
  {DeVore}}, \bibinfo {author} {\bibfnamefont {A.}~\bibnamefont {Gauthier}},
  \bibinfo {author} {\bibfnamefont {J.}~\bibnamefont {Levy}},\ and\ \bibinfo
  {author} {\bibfnamefont {C.}~\bibnamefont {Singh}},\ }\bibfield  {title}
  {\bibinfo {title} {Development and evaluation of a tutorial to improve
  students' understanding of a lock-in amplifier},\ }\href
  {https://doi.org/10.1103/PhysRevPhysEducRes.12.020127} {\bibfield  {journal}
  {\bibinfo  {journal} {Phys. Rev. Phys. Educ. Res.}\ }\textbf {\bibinfo
  {volume} {12}},\ \bibinfo {pages} {020127} (\bibinfo {year}
  {2016}{\natexlab{b}})}\BibitemShut {NoStop}%
\bibitem [{\citenamefont {Stanley}\ and\ \citenamefont
  {Lewandowski}(2018)}]{StanleyLabNotebooks}%
  \BibitemOpen
  \bibfield  {author} {\bibinfo {author} {\bibfnamefont {J.~T.}\ \bibnamefont
  {Stanley}}\ and\ \bibinfo {author} {\bibfnamefont {H.~J.}\ \bibnamefont
  {Lewandowski}},\ }\bibfield  {title} {\bibinfo {title} {Recommendations for
  the use of notebooks in upper-division physics lab courses},\ }\href
  {https://doi.org/10.1119/1.5001933} {\bibfield  {journal} {\bibinfo
  {journal} {American Journal of Physics}\ }\textbf {\bibinfo {volume} {86}},\
  \bibinfo {pages} {45} (\bibinfo {year} {2018})}\BibitemShut {NoStop}%
\bibitem [{\citenamefont {Dounas-Frazer}\ and\ \citenamefont
  {Lewandowski}(2017)}]{DounasFrazerInstructorTroubleshooting}%
  \BibitemOpen
  \bibfield  {author} {\bibinfo {author} {\bibfnamefont {D.~R.}\ \bibnamefont
  {Dounas-Frazer}}\ and\ \bibinfo {author} {\bibfnamefont {H.~J.}\ \bibnamefont
  {Lewandowski}},\ }\bibfield  {title} {\bibinfo {title} {Electronics lab
  instructors' approaches to troubleshooting instruction},\ }\href
  {https://doi.org/10.1103/PhysRevPhysEducRes.13.010102} {\bibfield  {journal}
  {\bibinfo  {journal} {Phys. Rev. Phys. Educ. Res.}\ }\textbf {\bibinfo
  {volume} {13}},\ \bibinfo {pages} {010102} (\bibinfo {year}
  {2017})}\BibitemShut {NoStop}%
\bibitem [{\citenamefont {Carlone}\ and\ \citenamefont
  {Johnson}(2007)}]{CarloneJohnsonIdentity}%
  \BibitemOpen
  \bibfield  {author} {\bibinfo {author} {\bibfnamefont {H.~B.}\ \bibnamefont
  {Carlone}}\ and\ \bibinfo {author} {\bibfnamefont {A.}~\bibnamefont
  {Johnson}},\ }\bibfield  {title} {\bibinfo {title} {Understanding the science
  experiences of successful women of color: Science identity as an analytic
  lens},\ }\href@noop {} {\bibfield  {journal} {\bibinfo  {journal} {Journal of
  Research in Science Teaching}\ }\textbf {\bibinfo {volume} {44}},\ \bibinfo
  {pages} {1187} (\bibinfo {year} {2007})}\BibitemShut {NoStop}%
\bibitem [{\citenamefont {Hazari}\ \emph {et~al.}(2013)\citenamefont {Hazari},
  \citenamefont {Sadler},\ and\ \citenamefont {Sonnert}}]{HazariIdentity}%
  \BibitemOpen
  \bibfield  {author} {\bibinfo {author} {\bibfnamefont {Z.}~\bibnamefont
  {Hazari}}, \bibinfo {author} {\bibfnamefont {P.~M.}\ \bibnamefont {Sadler}},\
  and\ \bibinfo {author} {\bibfnamefont {G.}~\bibnamefont {Sonnert}},\
  }\bibfield  {title} {\bibinfo {title} {The science identity of college
  students: Exploring the intersection of gender, race, and ethnicity},\ }\href
  {http://www.jstor.org/stable/43631586} {\bibfield  {journal} {\bibinfo
  {journal} {Journal of College Science Teaching}\ }\textbf {\bibinfo {volume}
  {42}},\ \bibinfo {pages} {82} (\bibinfo {year} {2013})}\BibitemShut {NoStop}%
\bibitem [{\citenamefont {Kalender}\ \emph
  {et~al.}(2019{\natexlab{a}})\citenamefont {Kalender}, \citenamefont
  {Marshman}, \citenamefont {Schunn}, \citenamefont {Nokes-Malach},\ and\
  \citenamefont {Singh}}]{KalenderIdentity1}%
  \BibitemOpen
  \bibfield  {author} {\bibinfo {author} {\bibfnamefont {Z.~Y.}\ \bibnamefont
  {Kalender}}, \bibinfo {author} {\bibfnamefont {E.}~\bibnamefont {Marshman}},
  \bibinfo {author} {\bibfnamefont {C.~D.}\ \bibnamefont {Schunn}}, \bibinfo
  {author} {\bibfnamefont {T.~J.}\ \bibnamefont {Nokes-Malach}},\ and\ \bibinfo
  {author} {\bibfnamefont {C.}~\bibnamefont {Singh}},\ }\bibfield  {title}
  {\bibinfo {title} {Why female science, technology, engineering, and
  mathematics majors do not identify with physics: They do not think others see
  them that way},\ }\href
  {https://doi.org/10.1103/PhysRevPhysEducRes.15.020148} {\bibfield  {journal}
  {\bibinfo  {journal} {Phys. Rev. Phys. Educ. Res.}\ }\textbf {\bibinfo
  {volume} {15}},\ \bibinfo {pages} {020148} (\bibinfo {year}
  {2019}{\natexlab{a}})}\BibitemShut {NoStop}%
\bibitem [{\citenamefont {Kalender}\ \emph
  {et~al.}(2019{\natexlab{b}})\citenamefont {Kalender}, \citenamefont
  {Marshman}, \citenamefont {Schunn}, \citenamefont {Nokes-Malach},\ and\
  \citenamefont {Singh}}]{KalenderIdentity2}%
  \BibitemOpen
  \bibfield  {author} {\bibinfo {author} {\bibfnamefont {Z.~Y.}\ \bibnamefont
  {Kalender}}, \bibinfo {author} {\bibfnamefont {E.}~\bibnamefont {Marshman}},
  \bibinfo {author} {\bibfnamefont {C.~D.}\ \bibnamefont {Schunn}}, \bibinfo
  {author} {\bibfnamefont {T.~J.}\ \bibnamefont {Nokes-Malach}},\ and\ \bibinfo
  {author} {\bibfnamefont {C.}~\bibnamefont {Singh}},\ }\bibfield  {title}
  {\bibinfo {title} {Gendered patterns in the construction of physics identity
  from motivational factors},\ }\href
  {https://doi.org/10.1103/PhysRevPhysEducRes.15.020119} {\bibfield  {journal}
  {\bibinfo  {journal} {Phys. Rev. Phys. Educ. Res.}\ }\textbf {\bibinfo
  {volume} {15}},\ \bibinfo {pages} {020119} (\bibinfo {year}
  {2019}{\natexlab{b}})}\BibitemShut {NoStop}%
\bibitem [{\citenamefont {Otero}\ and\ \citenamefont
  {Harlow}(2009)}]{OteroGettingStarted}%
  \BibitemOpen
  \bibfield  {author} {\bibinfo {author} {\bibfnamefont {V.~K.}\ \bibnamefont
  {Otero}}\ and\ \bibinfo {author} {\bibfnamefont {D.~B.}\ \bibnamefont
  {Harlow}},\ }\bibfield  {title} {\bibinfo {title} {Getting started in
  qualitative physics education research},\ }\href@noop {} {\bibfield
  {journal} {\bibinfo  {journal} {Reviews in PER Vol}\ }\textbf {\bibinfo
  {volume} {2}},\ \bibinfo {pages} {1} (\bibinfo {year} {2009})}\BibitemShut
  {NoStop}%
\bibitem [{\citenamefont {Wilcox}\ and\ \citenamefont
  {Lewandowski}(2016)}]{WilcoxOpenEnded}%
  \BibitemOpen
  \bibfield  {author} {\bibinfo {author} {\bibfnamefont {B.~R.}\ \bibnamefont
  {Wilcox}}\ and\ \bibinfo {author} {\bibfnamefont {H.~J.}\ \bibnamefont
  {Lewandowski}},\ }\bibfield  {title} {\bibinfo {title} {Open-ended versus
  guided laboratory activities: Impact on students' beliefs about experimental
  physics},\ }\href {https://doi.org/10.1103/PhysRevPhysEducRes.12.020132}
  {\bibfield  {journal} {\bibinfo  {journal} {Phys. Rev. Phys. Educ. Res.}\
  }\textbf {\bibinfo {volume} {12}},\ \bibinfo {pages} {020132} (\bibinfo
  {year} {2016})}\BibitemShut {NoStop}%
\bibitem [{\citenamefont {{Wilcox}}\ and\ \citenamefont
  {{Lewandowski}}(2016)}]{ECLASSValidation}%
  \BibitemOpen
  \bibfield  {author} {\bibinfo {author} {\bibfnamefont {B.}~\bibnamefont
  {{Wilcox}}}\ and\ \bibinfo {author} {\bibfnamefont {H.}~\bibnamefont
  {{Lewandowski}}},\ }\bibfield  {title} {\bibinfo {title} {Students'
  epistemologies about experimental physics: Validating the {Colorado Learning
  Attitudes about Science Survey} for experimental physics},\ }\href
  {https://doi.org/10.1103/PhysRevPhysEducRes.12.010123} {\bibfield  {journal}
  {\bibinfo  {journal} {Phys. Rev. Phys. Educ. Res.}\ }\textbf {\bibinfo
  {volume} {12}},\ \bibinfo {eid} {010123} (\bibinfo {year}
  {2016})}\BibitemShut {NoStop}%
\bibitem [{\citenamefont {Cohen}(2013)}]{CohenD}%
  \BibitemOpen
  \bibfield  {author} {\bibinfo {author} {\bibfnamefont {J.}~\bibnamefont
  {Cohen}},\ }\href@noop {} {\emph {\bibinfo {title} {Statistical power
  analysis for the behavioral sciences}}}\ (\bibinfo  {publisher} {Routledge},\
  \bibinfo {year} {2013})\BibitemShut {NoStop}%
\bibitem [{\citenamefont {Doucette}\ \emph {et~al.}(2020)\citenamefont
  {Doucette}, \citenamefont {Clark},\ and\ \citenamefont
  {Singh}}]{DoucetteHermione}%
  \BibitemOpen
  \bibfield  {author} {\bibinfo {author} {\bibfnamefont {D.}~\bibnamefont
  {Doucette}}, \bibinfo {author} {\bibfnamefont {R.}~\bibnamefont {Clark}},\
  and\ \bibinfo {author} {\bibfnamefont {C.}~\bibnamefont {Singh}},\ }\bibfield
   {title} {\bibinfo {title} {Hermione and the secretary: How gendered task
  division in introductory physics labs can disrupt equitable learning},\
  }\href {http://iopscience.iop.org/10.1088/1361-6404/ab7831} {\bibfield
  {journal} {\bibinfo  {journal} {European Journal of Physics}\ } (\bibinfo
  {year} {2020})}\BibitemShut {NoStop}%
\bibitem [{\citenamefont {Marshman}\ \emph {et~al.}(2018)\citenamefont
  {Marshman}, \citenamefont {Kalender}, \citenamefont {Nokes-Malach},
  \citenamefont {Schunn},\ and\ \citenamefont {Singh}}]{CauseForAlarm}%
  \BibitemOpen
  \bibfield  {author} {\bibinfo {author} {\bibfnamefont {E.~M.}\ \bibnamefont
  {Marshman}}, \bibinfo {author} {\bibfnamefont {Z.~Y.}\ \bibnamefont
  {Kalender}}, \bibinfo {author} {\bibfnamefont {T.}~\bibnamefont
  {Nokes-Malach}}, \bibinfo {author} {\bibfnamefont {C.}~\bibnamefont
  {Schunn}},\ and\ \bibinfo {author} {\bibfnamefont {C.}~\bibnamefont
  {Singh}},\ }\bibfield  {title} {\bibinfo {title} {Female students with {A}'s
  have similar physics self-efficacy as male students with {C}'s in
  introductory courses: A cause for alarm?},\ }\href@noop {} {\bibfield
  {journal} {\bibinfo  {journal} {Phys. Rev. Phys. Educ. Res.}\ }\textbf
  {\bibinfo {volume} {14}},\ \bibinfo {pages} {020123} (\bibinfo {year}
  {2018})}\BibitemShut {NoStop}%
\bibitem [{\citenamefont {Kalender}\ \emph {et~al.}(2020)\citenamefont
  {Kalender}, \citenamefont {Marshman}, \citenamefont {Schunn}, \citenamefont
  {Nokes-Malach},\ and\ \citenamefont {Singh}}]{KalenderSelfEfficacy}%
  \BibitemOpen
  \bibfield  {author} {\bibinfo {author} {\bibfnamefont {Z.~Y.}\ \bibnamefont
  {Kalender}}, \bibinfo {author} {\bibfnamefont {E.}~\bibnamefont {Marshman}},
  \bibinfo {author} {\bibfnamefont {C.~D.}\ \bibnamefont {Schunn}}, \bibinfo
  {author} {\bibfnamefont {T.~J.}\ \bibnamefont {Nokes-Malach}},\ and\ \bibinfo
  {author} {\bibfnamefont {C.}~\bibnamefont {Singh}},\ }\bibfield  {title}
  {\bibinfo {title} {Damage caused by women's lower self-efficacy on physics
  learning},\ }\href {https://doi.org/10.1103/PhysRevPhysEducRes.16.010118}
  {\bibfield  {journal} {\bibinfo  {journal} {Phys. Rev. Phys. Educ. Res.}\
  }\textbf {\bibinfo {volume} {16}},\ \bibinfo {pages} {010118} (\bibinfo
  {year} {2020})}\BibitemShut {NoStop}%
\bibitem [{\citenamefont {Singh}(2005)}]{SinghPeersTest}%
  \BibitemOpen
  \bibfield  {author} {\bibinfo {author} {\bibfnamefont {C.}~\bibnamefont
  {Singh}},\ }\bibfield  {title} {\bibinfo {title} {Impact of peer interaction
  on conceptual test performance},\ }\href@noop {} {\bibfield  {journal}
  {\bibinfo  {journal} {American journal of physics}\ }\textbf {\bibinfo
  {volume} {73}},\ \bibinfo {pages} {446} (\bibinfo {year} {2005})}\BibitemShut
  {NoStop}%
\bibitem [{\citenamefont {Azmitia}\ and\ \citenamefont
  {Montgomery}(1993)}]{AzmitiaFriends}%
  \BibitemOpen
  \bibfield  {author} {\bibinfo {author} {\bibfnamefont {M.}~\bibnamefont
  {Azmitia}}\ and\ \bibinfo {author} {\bibfnamefont {R.}~\bibnamefont
  {Montgomery}},\ }\bibfield  {title} {\bibinfo {title} {Friendship,
  transactive dialogues, and the development of scientific reasoning},\ }\href
  {https://doi.org/10.1111/j.1467-9507.1993.tb00014.x} {\bibfield  {journal}
  {\bibinfo  {journal} {Social Development}\ }\textbf {\bibinfo {volume} {2}},\
  \bibinfo {pages} {202} (\bibinfo {year} {1993})}\BibitemShut {NoStop}%
\bibitem [{\citenamefont {Heller}\ and\ \citenamefont
  {Heller}(2001)}]{HellerBook}%
  \BibitemOpen
  \bibfield  {author} {\bibinfo {author} {\bibfnamefont {P.}~\bibnamefont
  {Heller}}\ and\ \bibinfo {author} {\bibfnamefont {K.}~\bibnamefont
  {Heller}},\ }\href@noop {} {\emph {\bibinfo {title} {Cooperative Group
  Problem Solving in Physics}}}\ (\bibinfo  {publisher} {Brooks/Cole Publishing
  Company},\ \bibinfo {year} {2001})\BibitemShut {NoStop}%
\bibitem [{\citenamefont {Kirwan}\ and\ \citenamefont
  {Ainsworth}(1992)}]{KirwanTaskAnalysis}%
  \BibitemOpen
  \bibfield  {author} {\bibinfo {author} {\bibfnamefont {B.}~\bibnamefont
  {Kirwan}}\ and\ \bibinfo {author} {\bibfnamefont {L.~K.}\ \bibnamefont
  {Ainsworth}},\ }\href@noop {} {\emph {\bibinfo {title} {A Guide to Task
  Analysis: The Task Analysis Working Group}}}\ (\bibinfo  {publisher} {CRC
  Press},\ \bibinfo {year} {1992})\BibitemShut {NoStop}%
\bibitem [{\citenamefont {Clark}\ \emph {et~al.}(2007)\citenamefont {Clark},
  \citenamefont {Feldon}, \citenamefont {van Merriënboer}, \citenamefont
  {Yates},\ and\ \citenamefont {Early}}]{ClarkTaskAnalysis}%
  \BibitemOpen
  \bibfield  {author} {\bibinfo {author} {\bibfnamefont {R.~E.}\ \bibnamefont
  {Clark}}, \bibinfo {author} {\bibfnamefont {D.~F.}\ \bibnamefont {Feldon}},
  \bibinfo {author} {\bibfnamefont {J.~G.}\ \bibnamefont {van Merriënboer}},
  \bibinfo {author} {\bibfnamefont {K.}~\bibnamefont {Yates}},\ and\ \bibinfo
  {author} {\bibfnamefont {S.}~\bibnamefont {Early}},\ }\bibfield  {title}
  {\bibinfo {title} {Handbook of research on educational communications and
  technology},\ }in\ \href@noop {} {\emph {\bibinfo {booktitle} {Cognitive task
  analysis}}},\ \bibinfo {editor} {edited by\ \bibinfo {editor} {\bibfnamefont
  {J.~M.}\ \bibnamefont {Spector}}, \bibinfo {editor} {\bibfnamefont {M.~D.}\
  \bibnamefont {Merrill}}, \bibinfo {editor} {\bibfnamefont {J.~G.}\
  \bibnamefont {van Merriënboer}},\ and\ \bibinfo {editor} {\bibfnamefont
  {M.~P.}\ \bibnamefont {Driscoll}}}\ (\bibinfo  {publisher} {Lawrence Erlbaum
  Associates},\ \bibinfo {year} {2007})\BibitemShut {NoStop}%
\bibitem [{\citenamefont {Schwartz}\ \emph {et~al.}(2005)\citenamefont
  {Schwartz}, \citenamefont {Bransford},\ and\ \citenamefont
  {Sears}}]{BransfordSchwartz}%
  \BibitemOpen
  \bibfield  {author} {\bibinfo {author} {\bibfnamefont {D.~L.}\ \bibnamefont
  {Schwartz}}, \bibinfo {author} {\bibfnamefont {J.~D.}\ \bibnamefont
  {Bransford}},\ and\ \bibinfo {author} {\bibfnamefont {D.}~\bibnamefont
  {Sears}},\ }\bibfield  {title} {\bibinfo {title} {Efficiency and innovation
  in transfer},\ }in\ \href@noop {} {\emph {\bibinfo {booktitle} {Transfer of
  Learning: Research and Perspectives}}},\ \bibinfo {editor} {edited by\
  \bibinfo {editor} {\bibfnamefont {J.}~\bibnamefont {Mestre}}}\ (\bibinfo
  {publisher} {Information Age Publishing},\ \bibinfo {year} {2005})\ pp.\
  \bibinfo {pages} {1--51}\BibitemShut {NoStop}%
\bibitem [{\citenamefont {Kapur}(2008)}]{KapurProductiveStruggle}%
  \BibitemOpen
  \bibfield  {author} {\bibinfo {author} {\bibfnamefont {M.}~\bibnamefont
  {Kapur}},\ }\bibfield  {title} {\bibinfo {title} {Productive failure},\
  }\href@noop {} {\bibfield  {journal} {\bibinfo  {journal} {{Cognition and
  Instruction}}\ }\textbf {\bibinfo {volume} {26}},\ \bibinfo {pages} {379}
  (\bibinfo {year} {2008})}\BibitemShut {NoStop}%
\bibitem [{\citenamefont {Lorenzo}\ \emph {et~al.}(2006)\citenamefont
  {Lorenzo}, \citenamefont {Crouch},\ and\ \citenamefont
  {Mazur}}]{LorenzoCrouchMazur}%
  \BibitemOpen
  \bibfield  {author} {\bibinfo {author} {\bibfnamefont {M.}~\bibnamefont
  {Lorenzo}}, \bibinfo {author} {\bibfnamefont {C.~H.}\ \bibnamefont
  {Crouch}},\ and\ \bibinfo {author} {\bibfnamefont {E.}~\bibnamefont
  {Mazur}},\ }\bibfield  {title} {\bibinfo {title} {Reducing the gender gap in
  the physics classroom},\ }\href {https://doi.org/10.1119/1.2162549}
  {\bibfield  {journal} {\bibinfo  {journal} {American Journal of Physics}\
  }\textbf {\bibinfo {volume} {74}},\ \bibinfo {pages} {118} (\bibinfo {year}
  {2006})}\BibitemShut {NoStop}%
\bibitem [{\citenamefont {Karim}\ \emph {et~al.}(2018)\citenamefont {Karim},
  \citenamefont {Maries},\ and\ \citenamefont {Singh}}]{KarimEBAEgender}%
  \BibitemOpen
  \bibfield  {author} {\bibinfo {author} {\bibfnamefont {N.~I.}\ \bibnamefont
  {Karim}}, \bibinfo {author} {\bibfnamefont {A.}~\bibnamefont {Maries}},\ and\
  \bibinfo {author} {\bibfnamefont {C.}~\bibnamefont {Singh}},\ }\bibfield
  {title} {\bibinfo {title} {Do evidence-based active-engagement courses reduce
  the gender gap in introductory physics?},\ }\href@noop {} {\bibfield
  {journal} {\bibinfo  {journal} {Eur. J. Phys.}\ }\textbf {\bibinfo {volume}
  {39}},\ \bibinfo {pages} {025701} (\bibinfo {year} {2018})}\BibitemShut
  {NoStop}%
\end{thebibliography}%

\end{document}